\documentclass[a4paper,10pt]{article}
\usepackage[T1]{fontenc}
\usepackage[utf8]{inputenc}
\usepackage{lmodern}
\usepackage{amsmath,amssymb,graphicx,booktabs,bm,microtype}
\usepackage{fancyhdr,xcolor,caption,titlesec,placeins,float,needspace,etoolbox}
\usepackage{balance}
\usepackage[compress,nospace]{cite}
\usepackage[hidelinks,hypertexnames=false]{hyperref}
\newcommand{\ucite}[1]{$\!^\text{\cite{#1}}$}
\def\footnoterule{\kern 1mm \hrule width 7cm \kern 2.2mm}

\hypersetup{hidelinks,pdftitle={Hierarchical Fourier Phase Projection for Local Electronic Observables},
 pdfauthor={Tao Hu, Weiqing Zhou, Zhichang Fu, Yechen Chen, Shengjun Yuan}}
\titleformat{\section}{\normalfont\bfseries\large}{\thesection}{.6em}{}
\titleformat{\subsection}{\normalfont\bfseries}{\thesubsection}{.6em}{}
\titlespacing*{\section}{0pt}{11pt}{6pt}
\titlespacing*{\subsection}{0pt}{9pt}{5pt}
\begin{document}
\twocolumn[
\begin{center}
{\large\bfseries Hierarchical Fourier Phase Projection for Local Electronic Observables}\par
\vspace{7pt}
Tao Hu$^1$, Weiqing Zhou$^{1,2,*}$, Zhichang Fu$^1$, Yechen Chen$^1$, and Shengjun Yuan$^{1,2,3,\dagger}$\par
\vspace{4pt}
{\small
$^1$Key Laboratory of Artificial Micro- and Nano-Structures of Ministry of Education\\
and School of Physics, Wuhan University, Wuhan 430072, China\\
$^2$Wuhan Institute of Quantum Technology, Wuhan 430206, China\\
$^3$School of Artificial Intelligence, Wuhan University, Wuhan 430072, China}\par
\vspace{3pt}
{\footnotesize $^*$weiqingzhou@whu.edu.cn\qquad $^\dagger$s.yuan@whu.edu.cn}
\end{center}
\begin{center}
\begin{minipage}{.94\textwidth}
\small
In this letter, we develop hierarchical Fourier phase projection (HPP) for the calculation of local electronic observables. The method couples the spatial decay of the density matrix with a hierarchy of Fourier probes, improving accuracy through refinement, and making use of all previously evaluated responses. The variation in the calculated observables during the refinement process carries information on the residual error. We use a locality-based model to estimate the error from these changes with no extra probing calculations. In particular, we demonstrate the method for electron densities and nonlocal pseudopotential forces for fixed Kohn--Sham Hamiltonians of semiconductors, metals and thermally perturbed silica. The convergence is significantly accelerated by the spatial decay of the density matrix, especially if the density matrix is short-ranged. The number of probes necessary to achieve a specific local accuracy is only weakly dependent on the system size within the range of consideration. HPP provides a route toward linear-scaling electronic-structure calculations in which the computational effort for local observables can be matched to the required accuracy, offering an error-aware strategy for large-scale first-principles simulations.

\vspace{6pt}
\noindent\textbf{Keywords:} hierarchical probing; local electronic observables; electronic nearsightedness; linear-scaling electronic-structure methods; density matrix; Kohn--Sham density functional theory

\vspace{3pt}
\noindent\textbf{PACS:} 71.15.-m, 71.15.Dx, 02.70.-c
\end{minipage}
\end{center}
\vspace{6pt}
]
\label{start:letter}
Understanding the electronic properties of systems with defects, disorder, and interfaces requires accurate calculations on scales that remain challenging for conventional first-principles methods. A major drawback of widely-used Kohn--Sham approaches \ucite{ref1,ref2} is the reliance on explicitly resolving a large number of eigenstates, a process whose cubic asymptotic cost limits the achievable system sizes. Many electronic equilibrium observables can be evaluated from certain elements or contractions of the density matrix without the need to find individual eigenstates \ucite{ref5}. Electronic nearsightedness provides a physical basis for exploiting this distinction. Even if individual orbitals are extended, the density matrix can decay rapidly with increasing spatial separation, depending on the electronic structure and temperature \ucite{ref7,ref8,ref9}. This locality is the basis for linear scaling methods that employ spatial partitioning, localized orbitals, and direct approximations to the density matrix \ucite{ref3,ref4,ref6,ref10,ref11,ref12,ref13,ref14,ref15,ref16}.

The random-state approach calculates electronic observables via statistical sampling, instead of directly calculating individual eigenstates \ucite{ref17,ref18,ref19}. The spatial locality can be incorporated into the sampling vectors to improve the estimates. Structured probing cancels selected nearby cross terms rather than relying entirely on statistical averaging. Early work developed diagonal estimators using random or Hadamard vectors \ucite{ref20} and systematic probing schemes based on distance coloring \ucite{ref21}. Wang et al. established locality-enhanced convergence and devised gradient-based probing for local density matrix elements and forces\ucite{ref22}. The chromatic superposition states used in self-consistent density functional tight-binding calculations \ucite{ref23} are also derived from such coloring concepts. However, the probing distance required depends on the electronic structure and the desired accuracy and is thus usually unknown in advance \ucite{ref21,ref32}. A probing distance that is appropriate for one system may not be sufficient for another. If further refinement is needed, independently constructed colorings may not allow earlier operator responses to be reused \ucite{ref21,ref24}.

In a separate line of research, hierarchical probing addresses the reuse problem by organizing probe vectors into nested sequences. Originally designed for lattice quantum chromodynamics, it employs ordered Hadamard or Fourier vectors to increase the probing distance while retaining earlier operator responses \ucite{ref24}. Later work combines it with deflation \ucite{ref29} and extends it to lattices of arbitrary side lengths and general graphs \ucite{ref28}. These structures allow for stepwise improvement in a calculation, but do not by themselves quantify the remaining errors in the observable at each step. Separately, the differences in the initial probing results have been used to fit a decay model and to determine a probing distance in matrix entropy calculations \ucite{ref33}. A natural way forward for electronic observables is to link error estimation with reusable refinement via the spatial decay of the density matrix and the geometry of unresolved contributions.

We present hierarchical Fourier phase projection (HPP) for local electronic observables, which provides a unified computational framework for reusable hierarchical probing and residual estimation. Each refinement removes cross terms associated with a known set of grid-point pairs. We use the resulting changes in the observables to build an effective spatial decay model and then apply this model to the unresolved pairs to estimate the residual error. This framework considers the geometry of the eliminated and surviving pairs, rather than simply treating convergence as a function of the probing distance. The estimates do not require additional probing computations and the same operator responses can be reused for different levels of refinement and shared among different observables. We illustrate the technique using electron densities and nonlocal pseudopotential forces computed with fixed Kohn--Sham Hamiltonians for semiconductors, metals, and thermally perturbed silica. These studies quantify the relation between electronic locality and convergence, reveal a weak size dependence of the probe count required for a given local accuracy, and evaluate residual predictions for different ranges of electronic decay and cell geometries.

We write an operator $P$ in an orthonormal real-space basis $\{|i\rangle\}_{i=1}^{N_g}$ of $N_g$ grid points. The aim is to find observables of the form $O_\alpha=\operatorname{Re}\operatorname{Tr}(A_\alpha P)$, where $\alpha$ labels an observable component and $A_\alpha$ is the operator associated with $\alpha$. The identities below hold irrespective of spatial locality.

Each grid point is assigned a unique value $\Lambda_i\in\{0,\ldots,N_g-1\}$. At the refinement stage $L$, the number of logical probes $M_L$ is a factor of $N_g$. The probes and associated operator responses are
\begin{equation}
\begin{aligned}
|q_m^{(L)}\rangle&=\sum_{i=1}^{N_g}z_i\exp\!\left(\frac{2\pi\mathrm{i}m\Lambda_i}{M_L}\right)|i\rangle,\\
|y_m^{(L)}\rangle&=\widetilde P|q_m^{(L)}\rangle,
\end{aligned}
\label{eq:original2}
\end{equation}
with $m=0,\ldots,M_L-1$. Here, $\mathrm{i}$ is the imaginary unit and $\widetilde P$ is a fixed numerical approximation of $P$. The $z_i$ are independent Rademacher variables (symmetrically distributed over $\pm1$), sampled once and kept fixed during refinement, unless otherwise stated. The comparison with the constant selection $z_i=1$ is presented in the Supplementary Material (SM Sec.~S4.2). Each probe has unit-modulus components over the whole grid. Once the responses are available, each observable is estimated as
\begin{equation}
\widehat O_\alpha^{(L)}=\frac{1}{M_L}\sum_{m=0}^{M_L-1}\operatorname{Re}\langle q_m^{(L)}|A_\alpha|y_m^{(L)}\rangle.
\label{eq:original3}
\end{equation}
Let $a_L(i,j)$ be the coefficient left after factoring out the Rademacher signs. Fourier orthogonality gives
\begin{equation}
\begin{aligned}
a_L(i,j)&=\frac{1}{M_L}\sum_{m=0}^{M_L-1}e^{2\pi\mathrm{i}m(\Lambda_j-\Lambda_i)/M_L}\\
&=\begin{cases}1,&\Lambda_j\equiv\Lambda_i\pmod{M_L},\\0,&\text{otherwise}.\end{cases}
\end{aligned}
\label{eq:original4}
\end{equation}
At level $L$, we call two distinct grid points an unresolved pair if their labels have the same remainder modulo $M_L$. For a given point $i$, its alias class $\mathcal A_L(i)$ is the set of all points whose labels have the same remainder as $\Lambda_i$ modulo $M_L$.

\begin{figure}[t]
\centering
\includegraphics[width=\columnwidth]{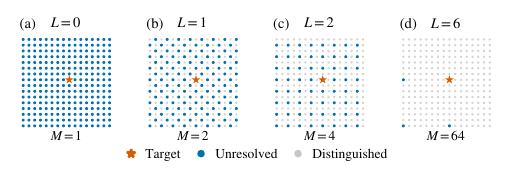}
\caption{Nested residual aliases on a $16\times16$ periodic grid for $L$ = 0, 1, 2, 6 with $M_L$ = 1, 2, 4, 64 probes. The orange star denotes the target grid point, blue points are still aliased, and gray points are distinguished. The minimum alias separations are $h$, $\sqrt{2}h$, $2h$, and $8h$ where $h$ is the grid spacing.}
\label{fig:original1}
\end{figure}

Following the coordinate construction of hierarchical probing \ucite{ref24,ref28}, we combine and order mixed-radix digits to distinguish nearby points early (SM Sec.~S1.2). The new digits of the label split the current alias classes, in the sense that $\mathcal A_{L+1}(i)\subseteq\mathcal A_L(i)$. The refinement shown in Fig.~\ref{fig:original1}(b) eliminates axial nearest neighbors, and subsequent levels progressively separate the remaining aliases. At level $L$, we define the minimum periodic distance between distinct aliased points as $d_{\mathrm{HPP}}(L)$. The blue dots in the illustration are aliased grid points. Each individual probe covers the whole grid.

Define $G_\alpha(i,j)=\langle i|A_\alpha\widetilde P|j\rangle$ and $\widetilde O_\alpha=\operatorname{Re}\operatorname{Tr}(A_\alpha\widetilde P)$. The projection error is then
\begin{equation}
\widehat O_\alpha^{(L)}-\widetilde O_\alpha=\operatorname{Re}\sum_i\sum_{\substack{j\in\mathcal A_L(i)\\j\ne i}}z_i z_jG_\alpha(i,j).
\label{eq:original5}
\end{equation}
Refinement removes unresolved pairs, so the minimum separation of the remaining pairs cannot decrease. The pair weight $K_{\alpha,ij}=\operatorname{Re}[G_\alpha(i,j)+G_\alpha(j,i)]$ is the coefficient of $z_i z_j$ for each pair $i<j$ after combining both directions in Eq.~(\ref{eq:original5}). The Rademacher mean-squared error is the sum of $K_{\alpha,ij}^2$ over the remaining pairs with $i<j$ (SM Sec.~S1.3). Refinement eliminates terms corresponding to nearby pairs from this nonnegative sum. Likewise, if the operator $A_\alpha\widetilde P$ is spatially local, the influence of distant pairs is also diminished.

When $M_{L+1}=bM_L$ for an integer $b$, earlier responses remain usable since $|q_m^{(L)}\rangle=|q_{bm}^{(L+1)}\rangle$. Only new probes require additional operator applications. When $M_L=N_g$, each alias class contains only a single point, so the projection error is zero regardless of the spatial structure of $\widetilde P$. The error due to the numerical approximation of $P$ remains separate. The full construction and derivation are presented in SM Sec.~S1.

For the electronic calculations, $P=f_\beta(H)$ is the density matrix per spin for a Kohn-Sham Hamiltonian $H$ \ucite{ref1}, where $f_\beta(E)$ is defined as $[1+\exp(\beta(E-\mu))]^{-1}$. In this context we define $\mu$ as the chemical potential and $\beta=(k_{\mathrm B}T)^{-1}$, where $T$ is the electronic temperature and $k_{\mathrm B}$ is Boltzmann's constant. For spin-unpolarized systems, the electron density and the nonlocal pseudopotential force are given by
\begin{equation}
\begin{aligned}
\rho_i&=\frac{2}{\Delta V}P_{ii},\\
F_{I\nu}^{\mathrm{NL}}&=-2\operatorname{Tr}\!\left(P\frac{\partial V_{\mathrm{NL}}}{\partial R_{I\nu}}\right),
\end{aligned}
\label{eq:original1}
\end{equation}
where $\Delta V$ is the volume of the grid cell, $V_{\mathrm{NL}}$ is the nonlocal pseudopotential, and $R_{I\nu}$ is the Cartesian position of atom $I$ in the direction $\nu$. The factor of two is due to spin degeneracy.

The density depends on the diagonal elements of $P$, while the nonlocal force also depends on the off-diagonal elements through spatially limited derivatives of the projectors. These two observables can be evaluated from the same responses, while other one-body observables can be obtained by changing $A_\alpha$. The spatial decay of the density matrix generally makes nearby unresolved pairs contribute more strongly, so resolving these pairs first reduces the dominant aliasing errors.

Calculations are performed with ABPLaS employing a real-space finite-difference Hamiltonian \ucite{ref19,1994-chelikowsky-higher}, the PBE functional \ucite{ref25}, and optimized norm-conserving Vanderbilt pseudopotentials \ucite{ref26}. Each hierarchy evaluates observables for a fixed Hamiltonian. The size series of C, Si, and Al extend along the $z$-axis with fixed cell lengths in the $x$ and $y$ directions, allowing long decay ranges to be examined at a manageable grid size. The electronic temperature is set to 300 K, unless otherwise stated. Reference observables are obtained from eigenstates of the same Hamiltonian by using scalar Fermi-Dirac occupations. All HPP trajectories used to test the predictor keep the Rademacher signs fixed on the physical grid. Density discrepancies are reported in percent as $100\|\widehat\rho-\rho^{\rm ref}\|_2/\|\rho^{\rm ref}\|_2$. The reference density norm is used only to convert absolute predictions and increments to percentages. The total atomic forces are computed by combining the nonlocal contribution with the local electron-ion, ion-ion and, if applicable, nonlinear core correction terms. The remaining force terms are evaluated from the electron density, ionic positions and prescribed core charge distributions. Force errors remain absolute RMS deviations over atomic Cartesian components, at a benchmark target of 10 meV/\AA{}. Computational details and numerical operator diagnostics are provided in SM Sec.~\ref{sec:smoriginal1}.

\begin{figure}[t]
\centering
\includegraphics[width=\columnwidth]{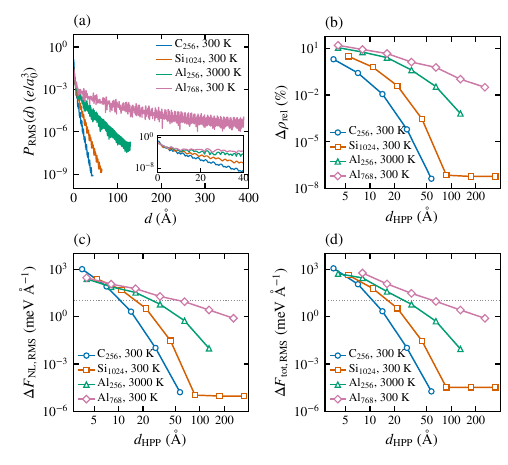}
\caption{Electronic locality and observable accuracy. (a) Amplitude of the density kernel as a function of the pair separation $d$ with an inset showing the near field. (b) Relative density RMS error and (c,d) nonlocal and total force RMS errors as a function of the minimum alias distance $d_{\mathrm{HPP}}$. The tails at the numerical floor are kept in these convergence paths. Total force RMS errors are calculated after summing the component error vectors. In (d), the cold Al $L$ = 9 point is missing, as the density estimate still has negative values at this coarse level, and the nonlinear core correction (NLCC) contribution was not computed. (c,d) The dotted lines are the 10 meV/\AA{} force target.}
\label{fig:original2}
\end{figure}

First we study how the spatial decay of the occupation kernel affects convergence. The RMS amplitude $P_{\mathrm{RMS}}(d)$ of $2\widetilde P_{ji}/\Delta V$ is plotted in Fig.~\ref{fig:original2}(a) for selected pairs of grid points as a function of the axial minimum-image separation $d$ (see SM Sec.~\ref{sec:sm-kernel-diagnostic}). C at 300 K shows the shortest range, followed by Si at 300 K and Al at 3000 K, whereas Al at 300 K exhibits an oscillatory tail over several hundred angstroms. The kernel diagnostic uses the pair separation $d$, while the HPP error curves use the minimum unresolved separation $d_{\mathrm{HPP}}$.

The convergence of density and nonlocal forces follows this sequence of ranges as shown in Figs.~\ref{fig:original2}(b,c). These results show that kernel decay strongly affects the convergence of both observables. The first sampled separations that satisfy the criteria for the nonlocal force are 14.2 and 21.7 \AA{} for C$_{256}$ and Si$_{1024}$ at 300 K, respectively, followed by 32.4 \AA{} for Al$_{256}$ at 3000 K and 64.7 \AA{} for Al$_{768}$ at 300 K. The C and Si errors are substantially reduced by later refinement before reaching the numerical error floor, whereas Al converges more slowly at low temperatures. These trends are related to the weights of the surviving pairs in equation (\ref{eq:original5}). Removing nearby cross terms is especially helpful for short range kernels, while for the density matrices with longer ranges, a larger distance between the remaining pairs is necessary.

\begin{figure}[t]
\centering
\includegraphics[width=\columnwidth]{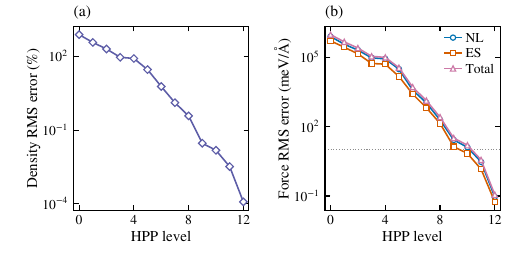}
\caption{Three-dimensional SiO$_2$ validation. (a) Relative density RMS error and (b) RMS errors of nonlocal, electrostatic and total forces. Total forces include NLCC, with component error vectors summed before RMS calculation. The 10 meV/\AA{} force target is represented by the dotted line in (b).}
\label{fig:sio2-observables}
\end{figure}

The respective total force errors are shown in Fig.~\ref{fig:original2}(d). The elongated periodic cells used here amplify the electrostatic force errors associated with residual density differences \ucite{tyagi2004}. We therefore correct the electrostatic term in this panel (SM Sec.~S5) before adding it to the nonlocal and NLCC forces. When the nonlocal force error first meets the target, the total force error is also below the target in all four cases. The three-dimensional silica test as shown in Fig.~\ref{fig:sio2-observables} supports this geometric explanation. In this test, the cell has similar dimensions along all three directions, and the uncorrected electrostatic force error remains smaller than the nonlocal force error during refinement.

\begin{figure}[t]
\centering
\includegraphics[width=\columnwidth]{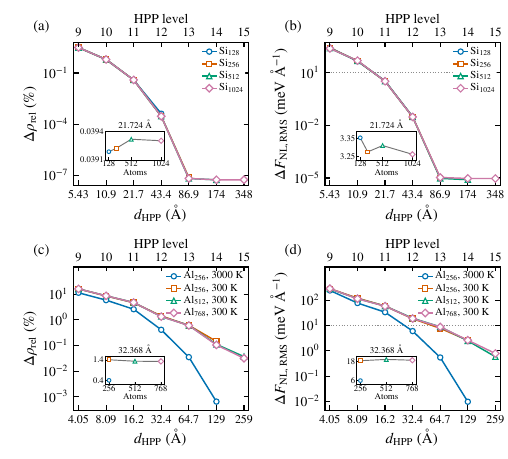}
\caption{Transferability of HPP accuracy across sizes. (a,b) Relative density RMS errors (\%) and nonlocal force RMS errors (meV/\AA{}) for silicon cells with 128--1024 atoms at 300 K, and (c,d) aluminum cells with 256--768 atoms at 300 K, and Al$_{256}$ at 3000 K. Each density curve is normalized by its own reference density RMS. The lower axes show $d_{\mathrm{HPP}}$, while the upper axes show the corresponding HPP levels. The dotted lines in (b,d) indicate the 10 meV/\AA{} force target. Levels $L$ = 9 and above are shown, including tails at the numerical floor but excluding endpoints at full decoding ($M_L = N_g$). The Si curves end at $L$ = 12, 13, 14, and 15 for 128, 256, 512, and 1024 atoms, respectively. Insets compare the same errors in terms of number of atoms at the same $d_{\mathrm{HPP}}$ (21.724 \AA{} for Si and 32.368 \AA{} for Al).}
\label{fig:original3}
\end{figure}

We then examine whether the same probing resolution maintains comparable errors in local observables as the system grows. Density and nonlocal force convergence for silicon cells containing 128--1024 atoms are shown in Figs.~\ref{fig:original3}(a,b). At $L$ = 11, all four sizes first reach the force target. This level uses $M_L=32{,}000$ probes and $d_{\mathrm{HPP}}=21.724$ \AA{}, with force errors of $\sim3$ meV/\AA{}. Thus the same absolute probe count is sufficient for an eightfold increase of the number of atoms.

The force convergence pattern is similar for all three Al sizes at 300 K, as shown in Figs.~\ref{fig:original3}(c,d). The force target is first reached at $L$ = 13, $M_L=128\,000$, and $d_{\mathrm{HPP}}=64.736$ \AA{} for the Al kernel, which has a much longer range. The 3000 K calculation for the same Al$_{256}$ cell reaches the force target one level earlier, with half the probe count and residual alias separation. The size and temperature trends link the required probing budget to the electronic decay range through the weights of surviving pairs. The size-independent probing budget supports linear-scaling HPP evaluation at fixed local accuracy. SM Sec.~\ref{sec:smoriginal2} has the relevant cost assessment and direct timings.

To verify whether the same convergence pattern is also applicable to cells with similar dimensions along all three axes, we investigate a thermally perturbed quartz (SiO$_2$) cell with 1350 atoms. With refinement, the errors in density and force decrease as shown in Fig.~\ref{fig:sio2-observables}. The relative density RMS error is 0.00326\% for 87,480 logical probes, and the RMS errors for nonlocal, electrostatic and total forces are 2.982, 1.446 and 3.631 meV/\AA{}. The structure and reference computation are given in SM Sec.~\ref{sec:sm-sio2}.

\begin{figure*}[t]
\centering
\includegraphics[width=\textwidth]{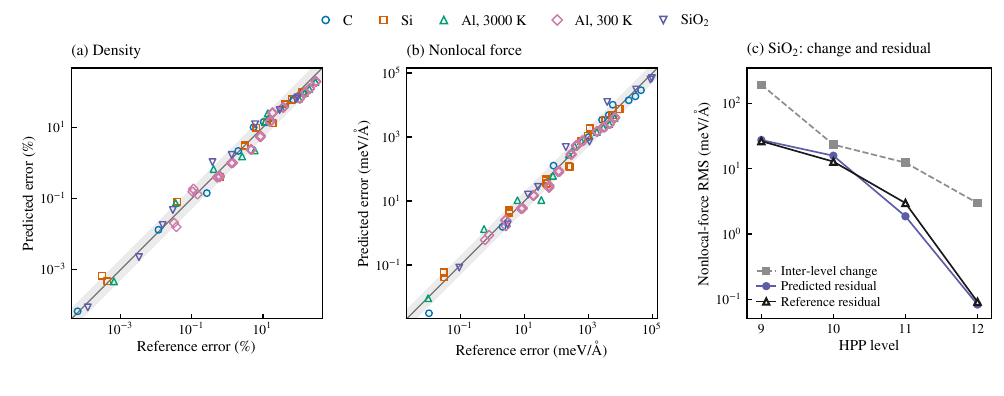}
\caption{Residual error estimates based on completed HPP refinements. Predicted versus reference errors for (a) density in relative RMS \% and (b) nonlocal force in meV/\AA{}. Each panel displays 143 checkpoints above the numerical error floor from 13 trajectories. The three Al$_{510}$ sign realizations are kept separate. Predicted and reference density errors use the same reference density norm. The diagonal line indicates perfect agreement between predicted and reference errors, while the gray region denotes deviations within a factor of two. (c) SiO$_2$ single-step increments, predicted and reference residuals, at $L$ = 9--12, showing the contributions removed by the last refinement versus those remaining. All forecasts are based only on completed increments.}
\label{fig:residual-prediction}
\end{figure*}

The same electronic locality that accelerates convergence also constrains the unresolved error. Prior work fitted a decay model to differences between initial probing results to select a graph-coloring distance for matrix entropy \ucite{ref33}. Here, the known geometry of discarded and remaining alias pairs links the completed changes to the remaining errors in local observables.

A completed refinement measures the impact of the pairs it discards. The remaining error is caused by the pairs it does not resolve. Let $\mathcal T_L$ be the remaining off-diagonal pairs at level $L$. An interval $[a,b]$ removes $\mathcal Q_{a,b}=\mathcal T_a\setminus\mathcal T_b$. Let $X$ denote the grid density or atomic Cartesian nonlocal force array, and $X^{(L)}$ its approximation at level $L$. We quantify the root-mean-square (RMS) endpoint difference $\eta_{X;a,b}=\operatorname{RMS}[X^{(b)}-X^{(a)}]$, with $a<b\le L$. The Rademacher average of $\eta_{X;a,b}^{2}$ equals the sum of squared weights over pairs $i<j$ in $\mathcal Q_{a,b}$, averaged over the components of $X$ (SM Sec.~\ref{sec:sm-sign-statistics}).

Electronic nearsightedness is reflected in the spatial decay of these pair weights, which also underlies bounds on probing errors \ucite{ref30}. Their squared magnitudes are modeled by the envelope $A_{X,L}r^{-3}e^{-2\kappa_{X,L}r}$, which is an algebraic prefactor multiplied by exponential attenuation. When $\kappa_{X,L}>0$, $\kappa_{X,L}^{-1}$ is an effective decay length for the pair amplitudes. For $\kappa_{X,L}=0$ there is an algebraic tail out to the full fitted range. The factor of two in the exponential indicates the squared magnitudes. $r^{-3}$ is motivated by the square of the $r^{-3/2}$ amplitude prefactor found for a three-dimensional tight-binding insulator \ucite{ref31}. We use it as a common physical prior, not a universal exponent for all materials or observables. The amplitude $A_{X,L}$ is positive and the inverse decay length $\kappa_{X,L}$ is nonnegative. Both are fitted separately for density and force at each level. For forces, the fitted envelope includes the weights of the nonlocal projector derivatives. For a family of pairs $\mathcal Q$, define
\begin{equation}
S_{\mathcal Q,L}(\kappa)=\sum_{0<r\le R_{\mathrm{cut},L}}n_{\mathcal Q}(r)r^{-3}e^{-2\kappa r},
\label{eq:pair-power}
\end{equation}
where $r$ is the periodic minimum-image distance and $n_{\mathcal Q}(r)$ counts the alias partners at separation $r$, averaged over target grid points. For both fitting and forecasting we use the same cutoff, $R_{\mathrm{cut},L}=d_{\mathrm{HPP}}(L+2)$ when aliases still persist at that level and the full periodic support otherwise. This cutoff depends only on the known phase geometry and does not require observable values from future refinement levels.

At each level, three completed endpoint differences are used to fit the amplitude and effective decay length by least-squares regression of the logarithmic RMS values. This balances relative discrepancies so that larger early changes do not dominate the fit solely through their magnitude. Taking the square roots of the sums of squared pair weights for the removed and remaining sets gives the modeled changes and residuals
\begin{equation}
\begin{aligned}
\widehat\eta_{X;a,b}&=\sqrt{A_{X,L}S_{\mathcal Q_{a,b},L}(\kappa_{X,L})},\\
\widehat\epsilon_{X,L}&=\sqrt{A_{X,L}S_{\mathcal T_L,L}(\kappa_{X,L})}.
\end{aligned}
\label{eq:residual-prediction}
\end{equation}
All systems use the same criterion for selecting fitting intervals, which depends only on pair geometry and the accuracy of previous one-step predictions (SM Sec.~\ref{sec:sm-prediction}). $\widehat\epsilon_{X,L}$ is an empirical estimate of the residual projection error for the fixed numerical operator. If the estimated residual exceeds the target, refinement can continue at the next level using all previously computed responses.

We test residual predictions over the same range of electronic decay and cell geometry using the same prediction rule for 13 trajectories on the semiconductor and metal size series, three Al$_{510}$ sign realizations and three-dimensional SiO$_2$. After removing seven checkpoints per observable dominated by the numerical error floor, 143 remain. Among these, 129 of the density forecasts and 130 of the nonlocal force forecasts fall within a factor of two of the reference error (see Figs.~\ref{fig:residual-prediction}(a, b)). The level sequences are given in SM Sec.~\ref{sec:sm-validation}.

Three independent Rademacher realizations of displaced Al$_{510}$ with two vacancies are fitted separately using their own completed changes. The ratios of predicted to reference errors are similar all along the refinement process without any averaging of observables or predictions (SM Sec.~\ref{sec:sm-modulation}).

The discrepancy between the residual error and the latest increment is shown in Fig.~\ref{fig:residual-prediction}(c). At SiO$_2$ level $L$ = 11 (87,480 probes), the nonlocal force increment per step is 12.417 meV/\AA{}, the predicted residual is 1.869 meV/\AA{}, and the reference error is 2.982 meV/\AA{}. Hence the last increment overestimates the residual error by a factor of more than four, while the locality-based assessment captures its smaller magnitude. When interpreting a finished refinement it is important to consider which pairs are left unresolved. These estimates can be refined by additional electronic response data if such data are readily available or inexpensive to acquire (SM Sec.~\ref{sec:sm-optional-response}).

The key conceptual point of HPP is that the residual error estimate of a local electronic observable becomes a computable quantity during refinement. Hierarchical Fourier probes remove cross terms associated with known sets of grid-point pairs, while the spatial decay of the density matrix suppresses the remaining distant contributions. Because the geometry of both the eliminated and the surviving pairs is known, the changes produced by completed refinements can be converted into an estimate of the remaining error without additional evaluations of matrix-vector products. Electronic nearsightedness is thus used not only to accelerate convergence but to quantify the accuracy attainable for the observable being calculated.

This turns the evaluation of local observables from a fixed-resolution probe into an accuracy-controlled refinement. The same operator responses are reused across refinement levels and shared among different observables. The present demonstration uses fixed Kohn–Sham Hamiltonians, and extending this error control to self-consistent iterations is a natural next step. More broadly, HPP provides a route toward linear-scaling first-principles calculations in which computational effort is allocated according to the accuracy required, rather than fixed in advance. This accuracy-aware refinement should benefit computational physics, materials science, and quantum chemistry, where large-scale first-principles calculations of local observables are often limited by the cost of resolving extended electronic states.

\paragraph*{Acknowledgments.}
This work is funded by the National Natural Science Foundation of China (Grant Nos. 12425407 and 12547165). W.Z. is grateful for the support from the Hubei Provincial Young Science and Technology Talent \textquotedblleft{}Morning Light\textquotedblright{} Support Program and the Hongyi postdoctoral fellowship at Wuhan University. We thank the Core Facility of Wuhan University for providing the computational resources.

\balance

\label{end:letter}

\clearpage
\nobalance
\onecolumn
\setcounter{page}{1}
\setlength{\emergencystretch}{3em}
\tolerance=200
\displaywidowpenalty=50
\AtBeginEnvironment{equation}{\setlength{\abovedisplayskip}{6pt}\setlength{\belowdisplayskip}{6pt}\setlength{\abovedisplayshortskip}{2pt}\setlength{\belowdisplayshortskip}{4pt}}
\setcounter{section}{0}\setcounter{subsection}{0}
\setcounter{equation}{0}\setcounter{figure}{0}\setcounter{table}{0}
\renewcommand{\thesection}{S\arabic{section}}
\renewcommand{\thesubsection}{\thesection.\arabic{subsection}}
\renewcommand{\theequation}{S\arabic{equation}}
\renewcommand{\theHequation}{sm.\arabic{equation}}
\renewcommand{\thefigure}{S\arabic{figure}}
\renewcommand{\theHfigure}{sm.\arabic{figure}}
\renewcommand{\thetable}{S\arabic{table}}
\renewcommand{\theHtable}{sm.\arabic{table}}
\begin{center}
{\Large\bfseries Supplementary Materials}\par
\vspace{5pt}
{\large Hierarchical Fourier Phase Projection for Local Electronic Observables}\par
\vspace{9pt}
Tao Hu$^1$, Weiqing Zhou$^{1,2,*}$, Zhichang Fu$^1$, Yechen Chen$^1$, and Shengjun Yuan$^{1,2,3,\dagger}$\par
\vspace{6pt}
{\small $^1$Key Laboratory of Artificial Micro- and Nano-Structures of Ministry of Education\\
and School of Physics, Wuhan University, Wuhan 430072, China\\
$^2$Wuhan Institute of Quantum Technology, Wuhan 430206, China\\
$^3$School of Artificial Intelligence, Wuhan University, Wuhan 430072, China\\
$^*$weiqingzhou@whu.edu.cn\quad $^\dagger$s.yuan@whu.edu.cn}
\end{center}
\label{start:sm}
\section{Projection identities and pair statistics}
\label{sec:sm-derivation}

\subsection{Phase averaging and errors in observables}

Let $P$ be the target operator and $\widetilde P$ its fixed numerical approximation. The identities below are true for the latter, regardless of its implementation. For $q_m^{(L)}(i)=z_i\exp(2\pi\mathrm{i}m\Lambda_i/M_L)$, with $z_i=\pm1$ fixed throughout refinement, define

\begin{equation}
C_L=\frac{1}{M_L}\sum_{m=0}^{M_L-1}|q_m^{(L)}\rangle\langle q_m^{(L)}|,
\qquad (C_L)_{ij}=z_i z_j a_L(i,j).
\label{eq:app-A1}
\end{equation}

At a completed level, Fourier orthogonality gives

\begin{equation}
a_L(i,j)=\frac{1}{M_L}\sum_{m=0}^{M_L-1}
 e^{2\pi\mathrm{i}m(\Lambda_j-\Lambda_i)/M_L}
=\begin{cases}1,&\Lambda_j\equiv\Lambda_i\pmod{M_L},\\0,&\text{otherwise}.\end{cases}
\label{eq:app-A2}
\end{equation}

This exact cancellation requires all probes of the level to be included. The sign in the exponent can be flipped without changing $a_L(i,j)$. Each probe component has unit modulus, so the diagonal entries of $C_L$ are all one.

Let $\mathcal A_L(i)=\{j:\Lambda_j\equiv\Lambda_i\pmod{M_L}\}$ and $G_\alpha(i,j)=\langle i|A_\alpha\widetilde P|j\rangle$. Then

\begin{equation}
\widehat O_\alpha^{(L)}=\operatorname{Re}\operatorname{Tr}(A_\alpha\widetilde P C_L)
=\operatorname{Re}\sum_{i,j}z_i z_j a_L(i,j)G_\alpha(i,j).
\label{eq:app-A3}
\end{equation}

The diagonal contribution $\widetilde O_\alpha=\operatorname{Re}\operatorname{Tr}(A_\alpha\widetilde P)$ can be subtracted, yielding

\begin{equation}
\begin{aligned}
E_\alpha^{(L)}&=\widehat O_\alpha^{(L)}-\widetilde O_\alpha
=\operatorname{Re}\operatorname{Tr}[A_\alpha\widetilde P(C_L-I)]\\
&=\operatorname{Re}\sum_i\sum_{\substack{j\in\mathcal A_L(i)\\j\ne i}}
 z_i z_jG_\alpha(i,j).
\end{aligned}
\label{eq:app-A4}
\end{equation}

Here $I$ is the identity matrix. Only unresolved off-diagonal pairs contribute. Also included in the total deviation from $O_\alpha=\operatorname{Re}\operatorname{Tr}(A_\alpha P)$ is the numerical operator error, $\widetilde O_\alpha-O_\alpha$.

\subsection{Nested spatial labels and reuse of responses}
\label{sec:sm-labels}

The construction for periodic lattices follows hierarchical probing \ucite{smref9,smref11}. Factor the number of grid points $N_\alpha$ on each axis into $T_\alpha$ integer radices,

\begin{equation}
N_\alpha=\prod_{s=1}^{T_\alpha}b_{\alpha s},\qquad\alpha=x,y,z.
\label{eq:app-A5}
\end{equation}

The $J=T_x+T_y+T_z$ mixed-radix coordinate digits are ordered and, where applicable, mixed in an invertible way into $c_1(i),\ldots,c_J(i)$, with radices $b_1,\ldots,b_J$. For example, two coordinate digits $d_1,d_2$ of the same scale and common radix $b$ can be transformed as

\begin{equation}
(d_1,d_2)\longmapsto((d_1+d_2)\bmod b,d_2).
\label{eq:app-A6}
\end{equation}

For $b=2$ this is $(d_1\oplus d_2,d_2)$. The labels and probe counts at completed levels are

\begin{equation}
\Lambda_i=\sum_{t=1}^{J}c_t(i)\prod_{u=1}^{t-1}b_u,
\qquad M_L=\prod_{t=1}^{L}b_t.
\label{eq:app-A7}
\end{equation}

When we reduce $\Lambda_i$ modulo $M_L$ we keep the first $L$ digits. Each further digit divides a class, so that $\mathcal A_{L+1}(i)\subseteq\mathcal A_L(i)$. In the 2D example, the pairs $(x_t\oplus y_t,y_t)$ ordered from fine to coarse scales produce a checkerboard pattern at the first level. For the $16\times16$ grid, $M_L=1,2,4,64$ give minimum separations $h,\sqrt2h,2h,8h$.

The spatial resolution is given by

\begin{equation}
d_{\rm HPP}(L)=\min_{\substack{i\ne j\\\Lambda_i\equiv\Lambda_j\ ({\rm mod}\ M_L)}}
 d_{\rm PBC}(\mathbf r_i,\mathbf r_j).
\label{eq:app-A8}
\end{equation}

Here $\mathbf r_i$ is the position of grid point $i$, and $d_{\rm PBC}$ is the periodic minimum image distance. The minimum distance cannot get smaller under nesting, though it may be equal at consecutive levels. Distance does not alone determine residual multiplicities or directions.

For $M_{L+1}=b_{L+1}M_L$,

\begin{equation}
|q_m^{(L)}\rangle=|q_{b_{L+1}m}^{(L+1)}\rangle,
\qquad |y_m^{(L)}\rangle=|y_{b_{L+1}m}^{(L+1)}\rangle.
\label{eq:app-A9}
\end{equation}

This means that each previous response can be reused for the same operator and modulation field. At full resolution, $M_L=N_g$, no distinct residual pairs remain, so that

\begin{equation}
C_{\rm full}=I,\qquad \widehat O_\alpha^{\rm full}=\widetilde O_\alpha.
\label{eq:app-A10}
\end{equation}

This removes projection error, no matter the locality, but leaves the numerical operator error.

When the operator and modulation are real, conjugate probes and their responses satisfy

\begin{equation}
|q_{M_L-m}^{(L)}\rangle=(|q_m^{(L)}\rangle)^*,\qquad
\widetilde P|q_{M_L-m}^{(L)}\rangle=(\widetilde P|q_m^{(L)}\rangle)^*.
\label{eq:app-A11}
\end{equation}

Indices are taken modulo $M_L$. If $A_\alpha$ is real, paired probes give the same real observable contribution. Thus, only $\lfloor M_L/2\rfloor+1$ independent inputs are needed, nearly halving the number of operator applications. Their multiplicities are retained, and normalization still uses the logical count $M_L$.

\subsection{Pair statistics and residual prediction}
\label{sec:sm-sign-statistics}

For each distinct pair $i<j$, combine both directions into $K_{\alpha,ij}=\operatorname{Re}[G_\alpha(i,j)+G_\alpha(j,i)]$. Each such unordered pair is counted once, and Eq.~\eqref{eq:app-A4} reads

\begin{equation}
E_\alpha^{(L)}=\sum_{i<j}a_L(i,j)z_i z_jK_{\alpha,ij}.
\label{eq:smoriginal5}
\end{equation}

The independence of the Rademacher signs ensures that $\mathbb E_z(z_i z_j)=0$ for $i\ne j$. In the squared sum, products from different unordered pairs also average to zero, even if the pairs share one index. Thus

\begin{equation}
\mathbb E_z E_\alpha^{(L)}=0,\qquad
\operatorname{Var}_z(E_\alpha^{(L)})=\sum_{i<j}a_L(i,j)K_{\alpha,ij}^{2}.
\label{eq:smoriginal6}
\end{equation}

Nesting removes terms from this nonnegative sum, so that the ensemble mean-square projection error cannot increase.

Let $\mathcal T_L$ be the surviving unordered pairs and $\mathcal Q_{a,b}=\mathcal T_a\setminus\mathcal T_b$ be the ones removed between completed levels $a<b$. For an observable array $X$ with $n_X$ components, collect its pair weights in $\mathbf K_{X,ij}$. The projection residual $\epsilon_{X,L}=\operatorname{RMS}[X^{(L)}-\widetilde X]$ and endpoint change $\eta_{X;a,b}=\operatorname{RMS}[X^{(b)}-X^{(a)}]$ satisfy

\begin{equation}
\begin{aligned}
\mathbb E_z\epsilon_{X,L}^{2}&=\frac{1}{n_X}\sum_{(i,j)\in\mathcal T_L}\|\mathbf K_{X,ij}\|_2^2,\\
\mathbb E_z\eta_{X;a,b}^{2}&=\frac{1}{n_X}\sum_{(i,j)\in\mathcal Q_{a,b}}\|\mathbf K_{X,ij}\|_2^2.
\end{aligned}
\label{eq:sm-prediction-identity}
\end{equation}

Here $\widetilde X$ is the result of full decoding of the same fixed numerical operator and the same sign field is used at both ends. For density, $n_X=N_g$, while for Cartesian nonlocal force, $n_X=3N_a$, where $N_a$ is the number of atoms. The force weights contain the nonlocal projector derivatives and need not be of the same spatial shape as the density weights.

The two sums show how refinement removes alias pairs while locality reduces the weight of distant pairs. The predictor from Section~\ref{sec:sm-prediction} uses completed changes to fit an effective radial envelope, then applies it to the remaining pair set.

\FloatBarrier

\section{Computational setup and numerical benchmarks}
\label{sec:smoriginal1}

\subsection{Fixed Hamiltonians and reference cells}

The calculations are carried out with the real-space finite-difference pseudopotential implementation in ABPLaS \ucite{smref1}, with

\begin{equation}
H[\rho_{\rm in}]=T_{\rm FD}+V_{\rm loc}+V_{\rm H}[\rho_{\rm in}]+V_{\rm xc}[\rho_{\rm in}]+V_{\rm NL}.
\label{eq:smoriginal1}
\end{equation}

Here $T_{\rm FD}$ is the finite-difference kinetic operator, $V_{\rm loc}$ and $V_{\rm NL}$ are the local and nonlocal pseudopotential terms and $V_{\rm H}$ and $V_{\rm xc}$ are the Hartree and exchange-correlation potentials calculated with the input density $\rho_{\rm in}$. In each hierarchy, the geometry, input density, effective potential, temperature and chemical potential are fixed. The tests evaluate observables without updating the Hamiltonian self-consistently. We employ PBE \ucite{smref4} with optimized norm-conserving Vanderbilt pseudopotentials \ucite{smref5} and the separable Kleinman-Bylander nonlocal term \ucite{smref6}. Electrostatics are treated with the periodic Poisson equation and Ewald splitting. The egg-box effect (dependence of pseudopotentials on grid position) is reduced by super-sampling followed by filtering onto the working grid \ucite{smref7}.

The regular cells are elongated along $z$ on $20\times20\times N_z$ grids (see Table~\ref{tab:smoriginal1}). Their finite difference stencil uses $N_h=6$ neighbors on each side of each Cartesian direction \ucite{smref2,smref3}. Electronic temperatures specify occupations, not thermal atomic disorder. The Al$_{510}$ modulation test adds two vacancies at asymmetric positions and a small displacement to an Al cell.

\begin{table}[htbp]
\centering
\caption{Benchmarks for regular cells. Grid spacings are approximate. Temperatures refer to electronic occupations.}
\label{tab:smoriginal1}
\begin{tabular}{llll}
\toprule
Material & Atom counts & Electronic temperature & Grid spacing (\AA)\\
\midrule
C & 128, 256, 512, 1024 & 300 K & 0.177\\
Si & 128, 256, 512, 1024 & 300 K & 0.271\\
Al & 256, 512, 768 & 300 K & 0.202\\
Al & 256 & 3000 K & 0.202\\
\bottomrule
\end{tabular}
\end{table}

\paragraph{Three-dimensional silica.}
\label{sec:sm-sio2}

For the quartz periodic cell, 450 Si and 900 O atoms are contained in a box of $25.5323\times24.5685\times27.0235$ \AA{}. The 300 K snapshot was generated with classical BKS dynamics \ucite{smref10}. Equilibration at fixed temperature and volume lasted 20 ps, followed by a 2 ps run at fixed energy. Fourfold Si and twofold O coordination are retained. The electronic calculation contains 7200 valence electrons, occupations at 300 K and a $108\times108\times120$ grid with spacings 0.23641, 0.22749 and 0.22520 \AA{}. The kinetic finite difference stencil involves six neighbors on either side of each Cartesian direction. The fixed Hamiltonian has a 7.2721 eV gap. The reference uses 3616 validated eigenstates with scalar Fermi-Dirac occupations. The fixed values are $\mu=0.13840198082023071$ Ha and $\beta=1052.5834160312877$ $\mathrm{Ha}^{-1}$.

\begin{figure}[htbp]
\centering
\includegraphics[width=70mm]{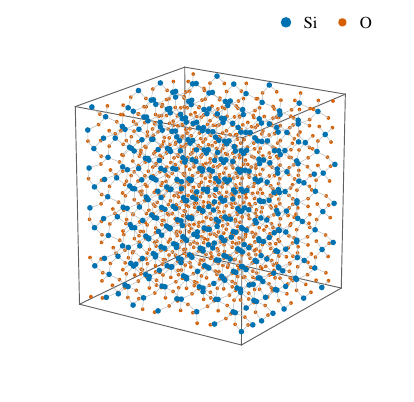}
\caption{Benchmark of thermally perturbed periodic quartz with 450 Si, 900 O atoms. The cell geometry and electronic calculation are reported above. The snapshot keeps the quartz coordination network.}
\label{fig:sm-sio2-structure}
\end{figure}

\Needspace{12\baselineskip}

\subsection{Conventions for density and force errors}

Let $\rho_{\rm ref,RMS}=\|\rho^{\rm ref}\|_2/\sqrt{N_g}$. Main text Eq.~\eqref{eq:residual-prediction} is computed in the original units of the observable. For benchmark reporting, we express the density convergence errors, predictions and changes in the same relative RMS percentage as

\begin{equation}
\begin{aligned}
\Delta\rho_{\rm rel}^{(L)}&=100\frac{\|\rho^{(L)}-\rho^{\rm ref}\|_2}{\|\rho^{\rm ref}\|_2},\\
\eta_{\rho;a,b}^{(\%)}&=\frac{100\eta_{\rho;a,b}}{\rho_{\rm ref,RMS}},\qquad
\widehat\epsilon_{\rho,L}^{(\%)}=\frac{100\widehat\epsilon_{\rho,L}}{\rho_{\rm ref,RMS}}.
\end{aligned}
\label{eq:sm-density-relative}
\end{equation}

For a given Hamiltonian and occupation the denominator is fixed. This is a global relative $L^2$ error, not the average of pointwise relative errors. The reference density norm is used only after the absolute prediction is made, and is not involved in its fit or window selection. Multiplying both prediction and reference residuals by the same factor does not change their ratio. No charge renormalization is employed.

Force errors are Cartesian component root-mean-square values,

\begin{equation}
\Delta F_{{\rm NL,RMS}}^{(L)}=
\left[\frac{1}{3N_a}\sum_{I,\nu}
 \left(F_{I\nu}^{{\rm NL},(L)}-F_{I\nu}^{{\rm NL,ref}}\right)^2\right]^{1/2},
\label{eq:smoriginal2}
\end{equation}

in meV/\AA{}, with a benchmark target of 10 meV/\AA{}. Force increments are in the same convention. Before taking the RMS, we sum the individual force contributions as

\begin{equation}
F_{I\nu}^{\rm tot}=F_{I\nu}^{\rm NL}+F_{I\nu}^{\rm es}+F_{I\nu}^{\rm NLCC}.
\label{eq:smoriginal3}
\end{equation}

Here $F^{\rm es}$ is the combined local electron--ion and ionic electrostatic term in the pseudocharge formulation as in the real-space force treatment of SPARC \ucite{smref12}. $F^{\rm NLCC}$ is the nonlinear core correction term.

\subsection{Locality of the kernel and numerical operator checks}
\label{sec:sm-kernel-diagnostic}

The kernel amplitude in main text Fig.~\ref{fig:original2}(a) is the RMS of $2\widetilde P_{ji}/\Delta V$ at 64 fixed representative grid origins $i$, with $j$ at the same transverse coordinates, and axial minimum-image separation $d$, $0\le d\le L_z/2$. $L_z$ is the axial cell length. This diagnostic describes the spatial decay of the fixed numerical occupation matrix. Its amplitude is given in units of $e/a_0^3$ and is distinct from the relative error of the full charge density field.

HPP probes and observable readouts still live on the full orthogonal real-space grid. For the four regular accuracy cases in main text Fig.~\ref{fig:original2}, the numerical occupation operators are constructed from 33-pole rational Fermi-Dirac actions \ucite{smref8}. Results at full decoding are compared with eigenstate references of the same Hamiltonian using scalar Fermi-Dirac occupations. Nonlocal force discrepancies and an independent check of selected matrix elements are shown in Table~\ref{tab:sm-operator-accuracy}. These checks include discrepancies in numerical representation, evaluation and reference.

\begin{table}[htbp]
\centering
\caption{Independent diagnostics of numerical operators. The column for full decoding gives RMS discrepancies of nonlocal force. The column for selected entries lists the largest absolute difference between $2\widetilde P_{ji}/\Delta V$ and its eigenstate benchmark over sampled pairs of source and target points $(i,j)$.}
\label{tab:sm-operator-accuracy}
\begin{tabular}{lrr}
\toprule
& Complete decoding & Selected kernel entries\\
System & NL force (meV/\AA) & Maximum ($e/a_0^3$)\\
\midrule
C$_{256}$ & $1.62575\times10^{-5}$ & $1.42\times10^{-10}$\\
Si$_{1024}$ & $9.32187\times10^{-6}$ & $2.53\times10^{-10}$\\
Al$_{256}$, 3000 K & $3.26313\times10^{-6}$ & $5.48\times10^{-11}$\\
Al$_{768}$, 300 K & $1.89808\times10^{-6}$ & $3.89\times10^{-11}$\\
\bottomrule
\end{tabular}
\end{table}

For SiO$_{2}$ the fixed numerical occupation operator is built from the reference eigenstates and scalar occupations. Its actions are then used to evaluate observables and changes between levels in later HPP calculations.

\Needspace{9\baselineskip}

\subsection{Direct HPP cost and timing}
\label{sec:smoriginal2}

The HPP cost of a sparse Hamiltonian is $O(pM_LN_g)$ where $p$ is the matrix-function work per grid point. At fixed accuracy, linear scaling requires both $p$ and the required probe count $M_L$ to remain bounded as $N_g$ increases. The weak size dependence of the required probe count in main text Fig.~\ref{fig:original3} supports the condition on $M_L$. Higher accuracy leads to a larger prefactor, whereas complete resolution, $M_L=N_g$, eliminates projection error at quadratic cost.

The C$_{128}$--C$_{1024}$ timings use a Chebyshev expansion of degree 2878 for an occupation smoothed with an error function of width 0.0248464 Ha. All four runs use four nodes with 64 cores each, the same batching and polynomial workload, and a total of 64 000 logical probes (32 001 independent conjugate representatives). Inherited responses are counted once. Timings include the accumulated direct HPP work. Operator preparation, residual prediction, other postprocessing, and self-consistency are not included in this measurement.

\begin{figure}[htbp]
\centering
\includegraphics[width=72mm]{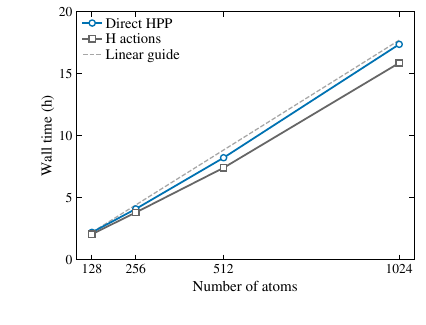}
\caption{Direct HPP cost at fixed probe load. Cumulative direct HPP time and the time spent applying the Hamiltonian for C$_{128}$--C$_{1024}$, using 64\,000 logical probes on four nodes with 64 cores each. The dashed line shows linear scaling normalized to the direct HPP time for C$_{128}$.}
\label{fig:timing}
\label{fig:smoriginal1}
\end{figure}

Figure~\ref{fig:timing} shows that cumulative HPP time and the time spent applying the Hamiltonian grow nearly linearly over this range of sizes at fixed probing workload.

\FloatBarrier

\section{Residual prediction from completed refinements}
\label{sec:sm-prediction}

\subsection{Periodic pair geometry and decay model}

The inputs at completed level $L$ are the density or nonlocal force arrays through level $L$ and the known labels assigned to the grid points. $n_Q(r)$ is the average number per target point of periodic alias partners at separation $r$, for each pair set $Q$. The histogram contains all relevant axial and diagonal displacements and their multiplicities. The wrapping uses the minimum image in all directions. A half-period shift is counted once. For SiO$_2$ at $L\ge9$, we average the pair distributions over 128 reference grid points with equal weights. Density and force are each fitted globally using these averaged pair distributions and the changes in the corresponding full arrays. Equation~\eqref{eq:sm-prediction-identity} combines the two directions of each pair before summing. The fixed difference in normalization relative to the displacement histograms is absorbed in the fitted amplitude.

The main text model uses squared weights $A_{X,L}r^{-3}\exp(-2\kappa_{X,L}r)$, where $A_{X,L}>0$ and $\kappa_{X,L}\ge0$. This algebraic factor is suggested by the $r^{-D/2}$ density-matrix prefactor derived for a tight-binding insulator in $D$ dimensions \ucite{smref13}. In three dimensions the amplitude prefactor is $r^{-3/2}$ and the squared prefactor is $r^{-3}$. We use this exponent as a common effective model for density and force in all the tested materials. Fixing it limits the tradeoff between algebraic and exponential decay when the fit is determined by only three aggregate observations. After completion of each level, the amplitude and effective inverse decay length are re-estimated. $\kappa_{X,L}=0$ corresponds to an algebraic regime over the distances observed.

Because of electronic locality, nearby unresolved pairs generally carry larger weights, which motivates a finite range for the geometric sums. The same cutoff $R_{{\rm cut},L}$ is used for all three fitting intervals, the residual set and the set removed at the next step. It is the nearest surviving alias distance at $L+2$. If there are no more aliases there, or this future geometry is not available, the full periodic support is preserved. The distances obey $r\le R_{{\rm cut},L}(1+10^{-13})$. This rule is based on known geometry and approximates the spatial weights of the full, untruncated endpoint changes. Extending the cutoff to the full periodic support gives similar results. The full periodic support gives 131/143 density predictions and 130/143 nonlocal force predictions within a factor of two, compared with 129/143 and 130/143 for the cutoff used in the main text. These checks support the common cutoff used in the main text results.

\subsection{Fitting the envelope to three observed changes}

We fit the envelope over a short history of refinement to describe the distances relevant to the current residual. Algebraic decay may dominate at shorter distances and the exponential damping may only be weakly resolved. A single fit to both early and late refinements does not necessarily represent the surviving tail. Three interval differences constrain the two parameters $A_{X,L}$ and $\kappa_{X,L}$ while allowing us to check the envelope's consistency and balance discrepancies among the observed changes.

For an interval $[a,b]$, $a<b\le L$, the input is the actual difference between the arrays, $\eta_{X;a,b}=\operatorname{RMS}[X^{(b)}-X^{(a)}]$. It is not obtained by adding scalar RMS values or their squares, as that would lose the correlations between increments. For each set of three fitting intervals, denote the measured endpoint changes as $D_j$ and the associated geometric sums from main text Eq.~\eqref{eq:pair-power} as $S_j(\kappa)$, $j=1,2,3$. The fit minimizes the equally weighted squared logarithmic RMS residuals. For fixed $\kappa$ we eliminate the amplitude analytically using

\begin{equation}
\begin{aligned}
b(\kappa)&=\frac13\sum_{j=1}^{3}\left[\log D_j-\frac12\log S_j(\kappa)\right],\\
J(\kappa)&=\frac13\sum_{j=1}^{3}\left[b(\kappa)+\frac12\log S_j(\kappa)-\log D_j\right]^2,
\qquad \log A=2b.
\end{aligned}
\label{eq:sm-log-fit}
\end{equation}

The logarithms compare multiplicative differences so that the largest early changes cannot dominate simply because of their absolute magnitude. The implementation minimizes $J$ over $q=\kappa a\in[0,20]$, where $a=20(h_x+h_y+h_z)/3$ and $h_x,h_y,h_z$ are the grid spacings. It evaluates zero, 96 logarithmically spaced values from $10^{-6}$ to 20, and 65 linearly spaced values from zero to 20, then applies bounded one-dimensional minimization at tolerance $10^{-12}$ to refine bracketed local minima, retaining the best boundary or interior result. Density and force are fitted separately with the same rule.

\subsection{Adaptive window selection for residual estimate}

Pairs may be removed in different directions and at different distances in successive refinements. These geometry changes are accommodated by a single adaptive rule that selects the three fitting intervals. Its recent candidate is the last three single-step intervals. Its grouped candidate is three adjacent, nonoverlapping intervals ending at $L$, each of length one, two or three for a total range of three to nine levels.

The choice of the grouped intervals is entirely geometrical. For each removed set and the current residual set, displacements within the common cutoff are assigned to 70 classes. The classes are defined by 7 combinations of nonzero axes $(x,y,z,xy,xz,yz,xyz)$, 2 half-period flags, and 5 radius bins relative to the minimum radius of that set, with boundaries $\sqrt2,2,4,8$. The weights $n(r)r^{-3}$ are normalized to distributions $u$. Of the at most $3^3=27$ interval combinations, we choose the one that minimizes the largest of the 6 squared Hellinger distances $\tfrac12\sum_c(\sqrt{u_c}-\sqrt{v_c})^2$ between the 4 distributions \ucite{smref14}. This favors similar pair distributions in the fitting intervals and the residual set. Candidate combinations are excluded if any fitting interval contains no removed pairs within the cutoff. Comparisons for radial binning use a relative tolerance of $10^{-12}$.

Each fit also predicts the next single-step change from $\mathcal T_L\setminus\mathcal T_{L+1}$ to compare the candidates without reference errors. At level $L$, these auxiliary forecasts are checked against the increments computed since the forecasts were issued. The score is the mean of $[\log(\widehat\eta_{X;j-1,j\mid j-1}/\eta_{X;j-1,j})]^2$ over the available $j=\max(4,L-2),\ldots,L$. Here $\widehat\eta_{X;j-1,j\mid j-1}$ is the change from level $j-1$ to $j$ predicted using data available at level $j-1$. A candidate requires at least two valid checks. Any zero or invalid prediction in this scoring window disqualifies that candidate for the current comparison. The candidate with the lower score is selected.

The final geometry check compares the current residual pair set to the last set that was removed. The same $n(r)r^{-3}$ weights, within the common cutoff, are classified into 14 classes by the combination of nonzero axes and the half-period flag. If the residual class with the highest weight is absent from the most recently removed set, we use the grouped window.

The chosen fit gives the present residual estimate in main text Eq.~\eqref{eq:residual-prediction}. The predicted change at the next level is later compared with the computed change to help select fitting intervals. At least three completed changes are required. Full decoding requires no extrapolation and has zero projection residual by construction. Selection or fitting does not include any future observable values or reference residuals.

\subsection{Evaluation and computational extent}

Prediction employs differences between endpoints, periodic geometric sums and small 1D fits, with no additional occupation operator applications. We test it by replaying existing HPP trajectories and using only the levels completed at each step.

The accompanying data contain interval end points, geometric histograms, fitted parameters, selected windows and validity flags. Reference residuals and density norms are stored separately for evaluation.

\FloatBarrier

\section{Prediction and modulation tests}
\label{sec:sm-validation}

\subsection{Residual calibration and evaluation set}

The 13 benchmark trajectories cover C$_{256}$, four Si sizes, Al$_{256}$ at 3000 K, three regular Al sizes at 300 K, three Al$_{510}$ sign realizations and 3D SiO$_{2}$. These trajectories cover system sizes and sign realizations in four families of materials. There are 150 predictions per observable before screening for the numerical floor. The main calibration statistics exclude seven checkpoints dominated by the numerical floor. These checkpoints are C$_{256}$ and Si$_{256}$ at $L=13$, Si$_{512}$ at $L=13,14$, and Si$_{1024}$ at $L=13,14,15$. The exclusion threshold is ten times the RMS difference between the fully decoded result and the reference result.

Of the remaining 143 checkpoints per observable, 129 density and 130 nonlocal force predictions are between one half and two times the reference residual. There are 4 density and 9 force predictions below half the reference residual, while 10 density and 4 force predictions exceed twice the reference residual. The ratios of predicted to reference residuals are 0.382--2.739 for density and 0.302--3.230 for force. Figure~\ref{fig:sm-prediction-trajectories} shows all prediction levels above the numerical floor for the ten material and size trajectories that are available. The three additional sign realizations are shown separately in Fig.~\ref{fig:sm-three-seed-prediction}(c, d).

\Needspace{11\baselineskip}

\subsection{Modulation and three sign realizations}
\label{sec:sm-modulation}

\begin{figure}[htbp]
\centering
\includegraphics[width=170mm]{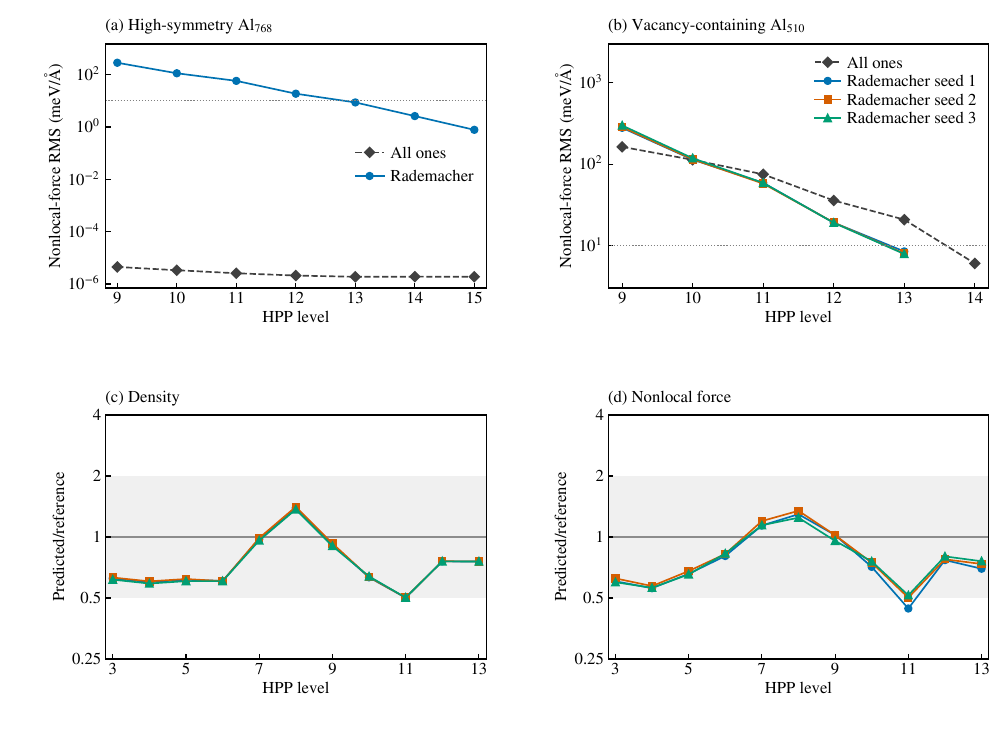}
\caption{Modulation and prediction among sign realizations at 300 K. (a,b) RMS errors of nonlocal forces for regular Al$_{768}$ and Al$_{510}$ with vacancies. Dotted lines are at 10 meV/\AA{}. (c,d) Ratios of predicted to reference residuals for density and nonlocal force for Al$_{510}$ over $L = 3\text{--}13$. Each sign field is fixed across levels and fitted separately using the common prediction rules. All three seeds are displayed separately.}
\label{fig:original4}
\label{fig:sm-three-seed-prediction}
\end{figure}

\begin{figure}[htbp]
\centering
\includegraphics[width=170mm]{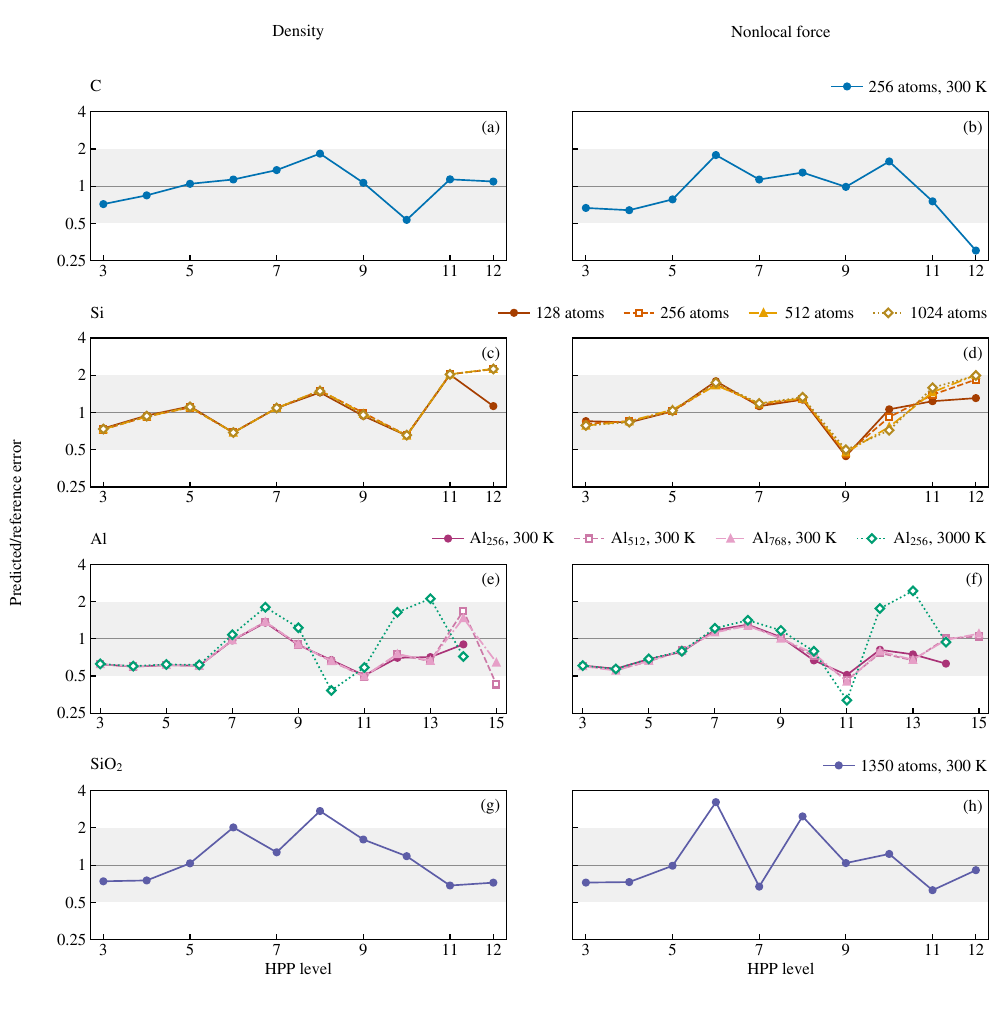}
\caption{Ratios of predicted to reference residuals for C, Si, Al and SiO$_{2}$. For the ten material and size trajectories, 110 checkpoints per observable are plotted, covering all available prediction levels above the numerical floor. Together with the 33 checkpoints from the three sign realizations in Fig.~\ref{fig:sm-three-seed-prediction}(c,d), these form the complete evaluation of 143 points for each observable. Unity indicates agreement, and the gray region indicates a factor of two.}
\label{fig:sm-prediction-trajectories}
\end{figure}

Figures~\ref{fig:original4}(a,b) vary the modulation but keep the same alias geometry. For Al$_{768}$, setting every sign to one gives much smaller nonlocal force errors, already below the target over the shown interval. This advantage is consistent with favorable cancellation in the force contraction for this highly symmetric structure. For Al$_{510}$ with vacancies the randomized sequences reach the target at 128 000 probes and $d_{\rm HPP}=64.736$ \AA{}. With every sign set to one, the error at this level is 20.78 meV/\AA{}, and it first passes at 256000 probes with 6.03 meV/\AA{}.

The three Al$_{510}$ sign fields are labeled seed 1, seed 2 and seed 3, corresponding to random number seeds 71041, 271828 and 314159 respectively. Each is fixed across refinement and provides its own completed changes for prediction. We keep all 11 available prediction levels, $L = 3\text{--}13$, separately for each seed, as shown in Figs.~\ref{fig:original4}(c,d). No averaging of observables or predictions over realizations is performed. The nonlocal force predictions at $L=13$ are 5.861--5.965 meV/\AA{}, while the reference residuals are 7.856--8.452 meV/\AA{}.

\Needspace{8\baselineskip}

\subsection{Three-dimensional alias geometry}
\label{sec:sm-sio2-validation}

The SiO$_{2}$ sequence is run on the physical grid with Rademacher seed 1 fixed. The labels assigned to the grid points determine the complete displacement histograms for all three periodic directions. The nearest alias distance alone does not provide the pair distribution information needed for prediction. The calculation at level $L = 12$ uses 174 960 logical probes and requires 87 481 independent inputs after conjugate reuse. Each alias class still contains eight grid points with nearest separation 13.51175 \AA{}. Thus small density and force errors are achieved before complete decoding. The figure data come with the full phase sequence. The evaluation covers all ten predictions of the current residual, $L = 3$ through 12.

\FloatBarrier

\clearpage

\section{Error prediction and electrostatic-force correction using source responses}
\label{sec:sm-optional-response}

The decay envelope of the predictor in Sec.~\ref{sec:sm-prediction} is fitted to completed HPP changes. If representative kernel responses are known in advance, or can be obtained at a small additional cost, their spatial detail can improve the estimated weights of the removed and surviving aliases. On the full real-space grid employed here, spatial correlations among responses from nearby sources allow representative samples to describe a much larger set of grid point pairs.

A unit source is one at grid point $s$ and zero elsewhere. Its response to $\widetilde P$ is the column $u_s(j)=\widetilde P_{js}$ of the numerical density matrix. For a pair set $\mathcal Q$, the response power $W_{\mathcal Q}$ is estimated by averaging over representative sources the sum of $|2u_s(j)/\Delta V|^2$ with the actual alias masks and the number of distinct periodic alias partners represented at each separation. Surviving and removed powers follow the same normalization. The rods use archived representative mean-square axial profiles after transverse decoding. Silica uses its three dimensional alias classes. For positive removed power and sufficient sampled support, the surviving set $\mathcal T_L$ and removed set $\mathcal Q_{L-1,L}$ give

\begin{equation}
\widehat\epsilon^{\rm resp}_{X,L}
=\eta_{X;L-1,L}
\sqrt{\frac{W_{\mathcal T_L}}{W_{\mathcal Q_{L-1,L}}}},
\label{eq:sm-response-prediction}
\end{equation}

where $\eta_{X;L-1,L}$ is the measured RMS density or nonlocal force change between the two finished levels. This construction requires one increment for calibration. The density kernel profile approximates the spatial shape of the projector-weighted response for force.

Graph coloring can combine several sources in one input when they are sufficiently far apart. With fixed local sampling density and source separation, the number of required inputs can be independent of the system size. The readout around each source can still contain tails of the other responses (cross-talk). Isolated-source controls can test whether the separation is sufficient.

For SiO$_{2}$ we divide the grid into 64 equal spatial regions and pick 88 sources in each of them, so we get 5632 sources. Periodic conflict coloring packs them into 3125 inputs, each with unit amplitudes at its chosen sites and zero elsewhere. 20.25084 \AA{} is the minimum separation between sources within an input. The 10 \AA{} readout is compared to 64 isolated-source controls to evaluate packing and truncation errors, and provides the response data for predictions at $L = 3\text{--}11$. The packed acquisition and these controls consume a total of 3189 operator inputs, or 7.3\% of the 43 741 independent HPP inputs accumulated through $L = 11$ after conjugate reuse. This comparison of input counts measures the extra response sampling.

\begin{figure}[htbp]
\centering
\includegraphics[width=170mm]{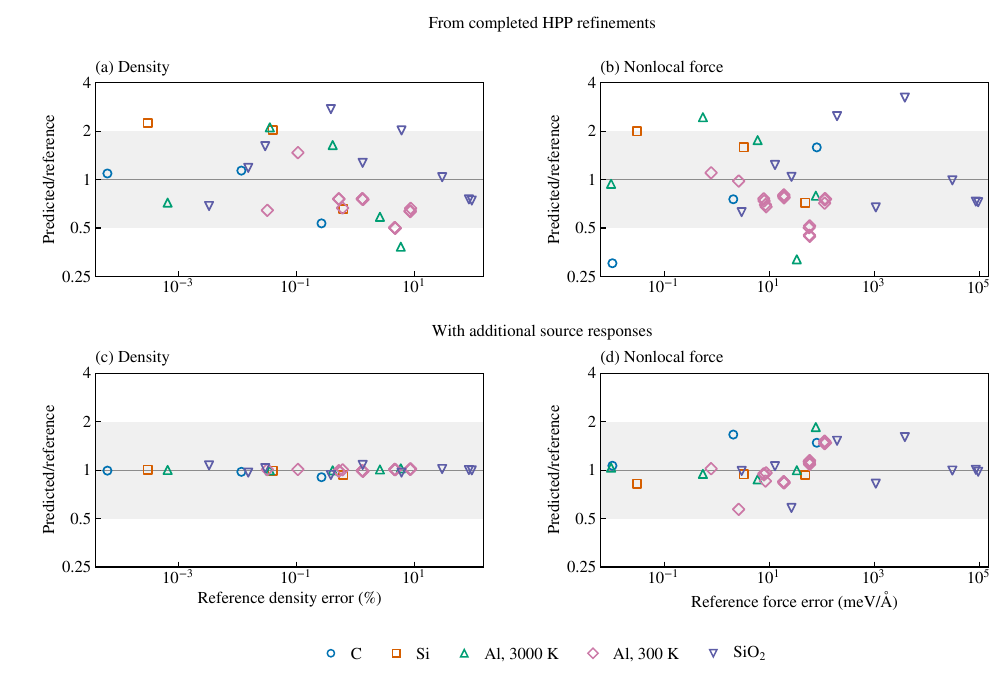}
\caption{Ratios of predicted to reference residuals at the same 38 checkpoints for each observable. Density is in the left column, nonlocal force in the right. The top row uses completed HPP changes with the main text $L+2$ cutoff. The bottom row adds source response information via Eq.~\eqref{eq:sm-response-prediction}. The reference error axis is identical for both rows in each column. Unity means agreement and the gray region indicates a factor of two.}
\label{fig:sm-optional-response}
\end{figure}

Figure~\ref{fig:sm-optional-response} compares the same 38 checkpoints per observable. These comprise 17 rod checkpoints, 12 combinations of seed and level for Al$_{510}$, and nine SiO$_{2}$ checkpoints. All 38 estimates using source responses fall within a factor of two, as compared to 31 for the main text model. Their ratios of predicted to reference residuals are 0.903--1.075 for density and 0.570--1.850 for nonlocal force. This extra spatial information thus improves the residual calibration on this common set, especially for the density.

The source responses obtained from the elongated rod cells for error prediction can be reused for electrostatic-force postprocessing. Using the response signs and phases, rather than simply their squared magnitudes, we can reconstruct and subtract selected residual alias contributions from the finite-level density before evaluating the electrostatic force. We correct the remaining alias contributions of the shortest axial displacements for $\mathrm{C}_{256}$, $\mathrm{Si}_{1024}$, and $\mathrm{Al}_{256}$ at $3000\,\mathrm{K}$, and the contributions at the three shortest displacements for $\mathrm{Al}_{768}$ at $300\,\mathrm{K}$, counting half-period partners only once. These settings are specific to the postprocessing of these benchmarks. Only the electrostatic component is replaced; the original nonlocal and NLCC forces are retained, and all components are combined as Cartesian vectors. The reported density and nonlocal-force benchmarks remain unchanged, and no such correction is applied to silica. Reusing available responses requires no additional operator applications. Reconstruction and electrostatic-force reevaluation incur separate costs.

\FloatBarrier
\Needspace{8\baselineskip}

\label{end:sm}


\begin{thebibliography}{34}
\footnotesize
\setlength{\itemsep}{0pt}
\setlength{\parskip}{0pt}

\bibitem{ref1}
Kohn W and Sham L J 1965 \textit{Phys. Rev.} \textbf{140} A1133

\bibitem{ref2}
Ratcliff L E, Mohr S, Huhs G, Deutsch T, Masella M and Genovese L 2017 \textit{WIREs Comput. Mol. Sci.} \textbf{7} e1290

\bibitem{ref5}
Dawson W, Degomme A, Stella M, Nakajima T, Ratcliff L E and Genovese L 2022 \textit{WIREs Comput. Mol. Sci.} \textbf{12} e1574

\bibitem{ref7}
Kohn W 1996 \textit{Phys. Rev. Lett.} \textbf{76} 3168

\bibitem{ref8}
Goedecker S 1998 \textit{Phys. Rev. B} \textbf{58} 3501

\bibitem{ref9}
Ismail-Beigi S and Arias T A 1999 \textit{Phys. Rev. Lett.} \textbf{82} 2127

\bibitem{ref3}
Goedecker S 1999 \textit{Rev. Mod. Phys.} \textbf{71} 1085

\bibitem{ref4}
Bowler D R and Miyazaki T 2012 \textit{Rep. Prog. Phys.} \textbf{75} 036503

\bibitem{ref6}
Zhou W, Fu Z and Yuan S 2026 \textit{Chem. Soc. Rev.} \textbf{55} 11494

\bibitem{ref10}
Yang W 1991 \textit{Phys. Rev. Lett.} \textbf{66} 1438

\bibitem{ref11}
Mauri F, Galli G and Car R 1993 \textit{Phys. Rev. B} \textbf{47} 9973

\bibitem{ref12}
Li X-P, Nunes R W and Vanderbilt D 1993 \textit{Phys. Rev. B} \textbf{47} 10891

\bibitem{ref13}
Goedecker S and Colombo L 1994 \textit{Phys. Rev. Lett.} \textbf{73} 122

\bibitem{ref14}
Goedecker S 1995 \textit{J. Comput. Phys.} \textbf{118} 261

\bibitem{ref15}
Palser A H R and Manolopoulos D E 1998 \textit{Phys. Rev. B} \textbf{58} 12704

\bibitem{ref16}
Niklasson A M N 2002 \textit{Phys. Rev. B} \textbf{66} 155115

\bibitem{ref17}
Yuan S, De Raedt H and Katsnelson M I 2010 \textit{Phys. Rev. B} \textbf{82} 115448

\bibitem{ref18}
Baer R, Neuhauser D and Rabani E 2013 \textit{Phys. Rev. Lett.} \textbf{111} 106402

\bibitem{ref19}
Zhou W and Yuan S 2023 \textit{Chin. Phys. Lett.} \textbf{40} 027101

\bibitem{ref20}
Bekas C, Kokiopoulou E and Saad Y 2007 \textit{Appl. Numer. Math.} \textbf{57} 1214

\bibitem{ref21}
Tang J M and Saad Y 2012 \textit{Numer. Linear Algebra Appl.} \textbf{19} 485

\bibitem{ref22}
Wang Z, Chern G-W, Batista C D and Barros K 2018 \textit{J. Chem. Phys.} \textbf{148} 094107

\bibitem{ref23}
Jiang Z, Xiao Z, Tang M, Li W, Sun Z, Xia K and Ke Y 2026 arXiv:\href{https://arxiv.org/abs/2605.20918}{2605.20918}

\bibitem{ref32}
Moussa J E and Baczewski A D 2019 \textit{Electron. Struct.} \textbf{1} 033001

\bibitem{ref24}
Stathopoulos A, Laeuchli J and Orginos K 2013 \textit{SIAM J. Sci. Comput.} \textbf{35} S299

\bibitem{ref29}
Gambhir A S, Stathopoulos A and Orginos K 2017 \textit{SIAM J. Sci. Comput.} \textbf{39} A532

\bibitem{ref28}
Laeuchli J and Stathopoulos A 2020 \textit{SIAM J. Sci. Comput.} \textbf{42} A1459

\bibitem{ref33}
Benzi M, Rinelli M and Simunec I 2023 \textit{Numer. Math.} \textbf{155} 377

\bibitem{1994-chelikowsky-higher}
Chelikowsky J R, Troullier N, Wu K and Saad Y 1994 \textit{Phys. Rev. B} \textbf{50} 11355

\bibitem{ref25}
Perdew J P, Burke K and Ernzerhof M 1996 \textit{Phys. Rev. Lett.} \textbf{77} 3865

\bibitem{ref26}
Hamann D R 2013 \textit{Phys. Rev. B} \textbf{88} 085117

\bibitem{tyagi2004}
Tyagi S 2004 \textit{Phys. Rev. E} \textbf{70} 066703

\bibitem{ref30}
Frommer A, Schimmel C and Schweitzer M 2021 \textit{SIAM J. Matrix Anal. Appl.} \textbf{42} 1290

\bibitem{ref31}
Taraskin S N, Drabold D A and Elliott S R 2002 \textit{Phys. Rev. Lett.} \textbf{88} 196405

\end{thebibliography}

\begin{thebibliography}{14}
\footnotesize
\setlength{\itemsep}{0pt}
\setlength{\parskip}{0pt}

\bibitem{smref9}
Stathopoulos A, Laeuchli J and Orginos K 2013 \textit{SIAM J. Sci. Comput.} \textbf{35} S299

\bibitem{smref11}
Laeuchli J and Stathopoulos A 2020 \textit{SIAM J. Sci. Comput.} \textbf{42} A1459

\bibitem{smref1}
Zhou W and Yuan S 2023 \textit{Chin. Phys. Lett.} \textbf{40} 027101

\bibitem{smref4}
Perdew J P, Burke K and Ernzerhof M 1996 \textit{Phys. Rev. Lett.} \textbf{77} 3865

\bibitem{smref5}
Hamann D R 2013 \textit{Phys. Rev. B} \textbf{88} 085117

\bibitem{smref6}
Kleinman L and Bylander D M 1982 \textit{Phys. Rev. Lett.} \textbf{48} 1425

\bibitem{smref7}
Ryu S, Choi S, Hong K and Kim W Y 2016 \textit{J. Chem. Phys.} \textbf{144} 094101

\bibitem{smref2}
Chelikowsky J R, Troullier N and Saad Y 1994 \textit{Phys. Rev. Lett.} \textbf{72} 1240

\bibitem{smref3}
Chelikowsky J R, Troullier N, Wu K and Saad Y 1994 \textit{Phys. Rev. B} \textbf{50} 11355

\bibitem{smref10}
van Beest B W H, Kramer G J and van Santen R A 1990 \textit{Phys. Rev. Lett.} \textbf{64} 1955

\bibitem{smref12}
Ghosh S and Suryanarayana P 2017 \textit{Comput. Phys. Commun.} \textbf{216} 109

\bibitem{smref8}
Lin L, Lu J, Ying L and E W 2009 \textit{Chin. Ann. Math. Ser. B} \textbf{30} 729

\bibitem{smref13}
Taraskin S N, Drabold D A and Elliott S R 2002 \textit{Phys. Rev. Lett.} \textbf{88} 196405

\bibitem{smref14}
Hellinger E 1909 \textit{J. reine angew. Math.} \textbf{136} 210

\end{thebibliography}
\end{document}